\documentclass[12pt, notitlepage,aps,pra,showpacs,superscriptaddress,
floatfix,nofootinbib,longbibliography]{revtex4-1}
\usepackage{amsmath,mathrsfs,amsbsy,color,graphicx,bm,amsthm,amsfonts}
\usepackage{times}
\usepackage[latin1]{inputenc}
\usepackage{graphicx}
\usepackage{physics}
\usepackage{slashed}
\usepackage{bbm}
\usepackage{subcaption}

\newtheorem{observation}{Observation}

\newtheorem{definition}{Definition}

\usepackage{cleveref}
\crefname{equation}{Eq.}{Eqs.}
\crefname{observation}{Obs.}{Obs.}
\crefname{lemma}{Lemma}{Lemmata}
\crefname{proof}{Proof}{Proofs}
\creflabelformat{proof}{#2proof#3}
\crefname{remark}{Remark}{Remarks}
\crefname{prop}{Proposition}{Propositions}

\begin{document}

\title{Implementation and representation of qudit multi-controlled unitaries and hypergraph states by N-body angular momentum couplings}

\author{F. E. S. Steinhoff} \email{steinhofffrank@gmail.com}
\affiliation{Universidade Federal de Mato Grosso, Faculdade de Ci\^ encia e Tecnologia, V\'arzea Grande, 78060-900, Brazil}
\affiliation{Centro Internacional de F\'isica, Instituto de F\'isica, Universidade de Bras\'ilia, 70910-900, Bras\'ilia, DF, Brazil}


\begin{abstract} 
We construct a representation of qudit multi-controlled unitary operators in terms of N-body angular momentum interactions. 
The representation is particularly convenient for odd-dimensional systems, with interesting connections to the Pegg-Barnett phase formalism. We illustrate the main points in the special case of qutrits, where simplifications and connections to dipole-quadrupole and quadrupole-quadrupole interactions can be established. We describe the representation of the closely related set of qudit hypergraph states, identifying possible realizations and their main obstacles. 
Qutrit tripartite controlled unitaries are decomposed in terms of more familiar two-body angular momentum couplings, 
enabling their implementation in a variety of physical systems. We give then a concrete example of implementation of qutrit unitaries and hypergraph states
in optical systems that employs single-photon sources, two-mode cross-Kerr interactions and linear optical operations. 
Moreover, we define a new set of states, called angular momentum hypergraph states, 
which are more directly related to the angular momentum representation.  
\end{abstract}
\maketitle

\section{Introduction}

Some of the most important concepts in quantum mechanics are expressed in terms of angular momentum interactions: the Ising and Heisenberg models \cite{huang}, spin-orbit coupling \cite{hamermesh}, the orbital angular momentum of light \cite{oam}, spin chirality \cite{xgwen1}, to name a few. From a fundamental point of view, the very division of particles into fermions or bosons depends on their intrinsic angular momentum, i.e. their spin, as well as the statistical behavior shown by each class \cite{spinstat}. From a practical point of view, angular momentum interactions are many times suited for experimental implementations of unitaries, states and measurements. 

In quantum information theory, most theoretical models and experiments employ series of two-body interactions between qubits, due to their relative mathematical simplicity and a greater familiarity in terms of practical realizations. However, a growing number of works have shown advantages in going beyond this paradigm, both by going to higher dimensions (qudits) \cite{campbell,nadish} or by using explicit N-body interactions \cite{3body,3spin0,3spin1,3spin2,5body}.


The present work shows a straightforward connection between qudit multi-controlled unitaries and angular momentum N-body couplings, 
especially in the case of odd-dimensional systems. This connection is convenient as a representational tool, 
enabling the identification of relevant interactions and the characterization of properties such as symmetries and correlations, 
as well as bringing in novel mathematical manipulations. 
Physically, our findings highlight the importance of synthesizing the N-body interactions described throughout the text, 
which may result in cheaper and more direct alternative implementations of multi-controlled unitaries. 
Moreover, we build upon previous results on the decomposition of multi-controlled unitaries in terms of local gates and bipartite controlled ones, in order to bring the constructions closer to technology available currently.

Parallel to our findings is the representation and implementation of qudit hypergraph states in terms of angular momentum interactions. 
Qudit hypergraph states \cite{qudithypergraph1,qudithypergraph2} are the generalization to higher-dimensional systems of the class of multiqubit hypergraph states \cite{qubithypergraph1,qubithypergraph2,qubithypergraph3}, 
which have attracted growing interest in recent years, with a recent experimental realization in silicon photonics \cite{exphypergraph}; 
previously, certain subsets of hypergraph states were identified in \cite{byoshida,miyake}. 
We believe our approach has a broader scope, with both a mathematical appeal and possibilities of physical realizations in quantum N-body systems.

A quantum optical analog of the angular momentum representation is obtained via a multi-mode version of the Jordan-Schwinger map \cite{multijsm}. 
As a proof of concept, we give a multi-rail implementation of qutrit multi-controlled gates and hypergraph states which relies solely on single-photon sources, two-mode cross-Kerr interactions, and linear optical elements, resources currently available in many optical setups. 
In principle, this scheme can be extended to any odd-dimensional system; however, this comes at the expense of a considerable increase in the number of modes. 

In order to circumvent some limitations of qudit hypergraph states, we define a similar class of states termed angular momentum hypergraph states, which can be represented solely in terms of angular momentum and are more directly related to quantum N-body systems. 
Our goal is to prepare the terrain for future use of angular momentum hypergraph states in condensed matter or quantum optical network scenarios. 

The paper is organized as follows: in Section II, we give a brief review of the angular momentum representation, qudit quantum gates and qudit hypergraph states, as well as setting some notation and conventions. In Section III, we establish a representation of qudit gates and hypergraph states in terms of angular momentum interactions, showing connections to the Pegg-Barnett phase formalism, which is briefly explained. The Clifford set of unitaries is expressed in terms of suitable N-body angular momentum interactions and possibilities for the realization of these interactions are suggested based on results in the literature. The angular momentum representation of qudit hypergraph states is established and we identify the main obstacles for practical implementations of this set of states. Moreover, we use previous results on qutrit circuit equivalences to decompose tripartite qutrit controlled gates in terms of two-body and local ones. In Section IV, we give a quantum optical analogue of the angular momentum representation, obtaining concrete setups for the implementation of qutrit gates of interest and three-qutrit hypergraph states. In Section V, we give the definition and main properties of angular momentum hypergraph states, a modification of the set of qudit hypergraph states that is expressed entirely in terms of the angular momentum operators and their couplings. Finally, in Section VI we give conclusions and perspectives on future developments.

\section{Preliminaries}

For simplicity, in what follows we set $\hbar=1$. We consider composite state spaces
$\mathcal{H}=\mathcal{H}_1\otimes\mathcal{H}_2\otimes\ldots\otimes\mathcal{H}_n$; without loss of generality, 
we take subsystems of equal dimensionalities, that is, $dim(\mathcal{H}_{\nu})=d$, $\nu=1,2,\ldots, n$. 
We adopt the notation $|\psi,\phi\rangle=|\psi\rangle\otimes|\phi\rangle$ and similarly write $A\otimes B=AB$ from Section V onwards. 
We express an unitary as $U=\exp(i\theta H)$ with a positive argument for mathematical convenience; 
obviously, one can switch to a negative argument convention by subtracting $2\pi$ of the corresponding phase.


\subsection{Angular momentum representation}

Hermitean operators $J_x$, $J_y$ and $J_z$ satisfying the commutation relations 
\begin{eqnarray}
[J_x,J_y]=iJ_z, \ \ \ [J_y,J_z]=iJ_x ,\ \ \ [J_z,J_x]=iJ_y,
\end{eqnarray}
are called {\it angular momentum operators} \cite{sakurai} and span the Lie-algebra $su(2)$; these operators can be seen as the components of an angular momentum vector operator $\mathbf{J}=(J_x,J_y,J_z)$. The corresponding Lie group $SU(2)$ is the set of unitaries $U(\phi_x,\phi_y,\phi_z)=\exp[i(\phi_xJ_x+\phi_yJ_y+\phi_zJ_z)]$, with $\phi_x,\phi_y,\phi_z\in\mathbb{R}$;
in particular, we define unitaries $R_l(\theta)=\exp(i\theta J_l)$, which represent rotations of an angle $\theta$ around the axis $l=x,y,z$. 
The Casimir operator $J^2=J_x^2+J_y^2+J_z^2$ is a $SU(2)$ invariant, physically representing the modulus squared of the total angular momentum. It commutes with the components $J_l$, $l=x,y,z$. An orthonormal basis of the state space is formed by the set $\{|j;m\rangle\}_{m=-j}^{m=j}$ of simultaneous eigenstates of the commuting observables $J^2|j;m\rangle = j(j+1)|j;m\rangle$ and $J_z|j;m\rangle=m|j;m\rangle$, where $j\in\{0,1/2,1,3/2,2,\ldots\}$ and $m\in\{j,j-1,\ldots,-j+1,-j\}$. The state space $\mathcal{H}$ admits then the direct-sum decomposition 
$\mathcal{H}=\bigoplus_j\mathcal{H}_j$ and is customary to work with a fixed value of $j$, i.e., within $\mathcal{H}_j$.

For low values of $j$ it is sometimes interesting to consider the matrix $[M]_j$ that represents a given operator $M$ in the ordered basis 
$\{|j;m=j\rangle,|j;m=j-1\rangle,\ldots,|j;m=-j\rangle\}$ of $\mathcal{H}_j$.
For example, for $j=1/2$ the components of $\mathbf{J}$ correspond to the well-known Pauli matrices (multiplied by $1/2$), while for $j=1$ we have  
\begin{eqnarray}
[J_x]_j= \frac{1}{\sqrt{2}}\left(\begin{array}{ccc}
{0}&{1}&{0}\\
{1}&{0}&{1}\\
{0}&{1}&{0}
\end{array}\right); \ \
[J_y]_j= \frac{1}{\sqrt{2}}\left(\begin{array}{ccc}
{0}&{-i}&{0}\\
{i}&{0}&{-i}\\
{0}&{i}&{0}
\end{array}\right); \ \ 
[J_z]_j= \left(\begin{array}{ccc}
{1}&{0}&{0}\\
{0}&{0}&{0}\\
{0}&{0}&{-1}
\end{array}\right). 
\end{eqnarray}

An important realization of the angular momentum operators is obtained through the two-mode {\it Jordan-Schwinger map} \cite{schwinger,puri}, 
\begin{eqnarray}
J_x=\frac{1}{2}(a^{\dagger}b+ab^{\dagger}), \ \ \ J_y=\frac{1}{2i}(a^{\dagger}b-ab^{\dagger}), \ \ \ J_z=\frac{1}{2}(a^{\dagger}a-b^{\dagger}b),
\end{eqnarray}
where $a$, $a^{\dagger}$, $b$ and $b^{\dagger}$ are two-mode bosonic operators satisfying the canonical commutation relations
\begin{eqnarray}
[a,a^{\dagger}]=[b,b^{\dagger}]=I; \ \ [a,b]=[a^{\dagger},b^{\dagger}]=[a,b^{\dagger}]=[a^{\dagger},b]=0, 
\end{eqnarray}
with $I$ representing the identity operator. An easy calculation shows that $J^2=\frac{N}{2}\left(\frac{N}{2}+I\right)$, where $N=a^{\dagger}a+b^{\dagger}b$ is the total number of particles. If $n_a$ and $n_b$ are the respective eigenvalues of $a^{\dagger}a$ and $b^{\dagger}b$, we can connect the quantum numbers through $j=(n_a+n_b)/2$ and $m=(n_a-n_b)/2$. 
Beyond its connection to quantum interferometry \cite{interf,yurke} and quantum metrology \cite{su2su11}, the Jordan-Schwinger correspondence is very useful for calculations involving the matrix elements of $[J_k]_j$. A recent multi-mode generalization of the Jordan-Schwinger map developed in \cite{multijsm} allows us to devise a quantum optical analog of the N-body angular momentum representation obtained. 

\subsection{Local and multi-controlled unitaries}

\subsubsection{Local Pauli and Clifford groups}

Let $\mathcal{H}$ be a $d-$dimensional system with orthonormal basis $\{|q\rangle\}_{q=0}^{d-1}$ and define the unitaries
\begin{eqnarray}
Z=\sum_{q=0}^{d-1}\omega^q|q\rangle\langle q|, \ \ \ X=\sum_{q=0}^{d-1}|q+1\rangle\langle q|,
\end{eqnarray}
where $\omega=e^{i2\pi/d}$ is the $d-$th root of unity and arithmetic operations are modulo $d$. 
These gates are related through the discrete Fourier transform (DFT) $F=d^{-1/2}\sum_{q,q'=0}^{d-1}\omega^{qq'}|q'\rangle\langle q|$ via
$X=FZF^{\dagger}$. 
With the properties $Z^d=X^d=I$ and $X^aZ^b=\omega^{-ab}Z^bX^a$, for $a,b\in\mathbb{Z}_d$, 
the unitaries $X^aZ^b$ span the $d-$dimensional {\it local Pauli group}. 

The local Pauli group is brought into itself under group conjugation by the following unitaries
\begin{eqnarray}
S(\xi,0,0)=\sum_{q=0}^{d-1}|\xi q\rangle\langle q|,\ \ \ S(1,\xi,0)=\sum_{q=0}^{d-1}\omega^{\xi q^2/2}|q\rangle\langle q|, \ \ \
S(1,0,\xi)=\sum_{q=0}^{d-1}\omega^{-\xi q^2/2}|p_q\rangle\langle p_q|,
\end{eqnarray}
where $|p_q\rangle=F|q\rangle$ are the eigenstates of $X$; in particular, 
we denote $|+\rangle=|p_0\rangle=1/\sqrt{d}(|0\rangle+|1\rangle+\ldots+|d-1\rangle)$. 
These together with the unitaries from the local Pauli group span the 
{\it local Clifford group} in $d$ dimensions, 
which is the normalizer of the Pauli group when considered as a subgroup of the full unitary group $U(d)$. 
When $d$ is a power of a prime number, we can picture the operations of the local Clifford group as transformations of a discrete phase-space \cite{vourdas}. 

We denote by $[M]_c$ the matrix representation of an operator $M$ in the computational basis $\{|q\rangle\}_{q=0}^{d-1}$. 
For example, in $d=3$ we have 
\begin{eqnarray*} 
[Z]_c=\left(\begin{array}{ccc}
{1}&{0}&{0}\\
{0}&{\omega}&{0}\\
{0}&{0}&{\omega^2}
\end{array}\right); \ \ \ [X]_c=\left(\begin{array}{ccc}
{0}&{0}&{1}\\
{1}&{0}&{0}\\
{0}&{1}&{0}
\end{array}\right); \ \ \ [F]_c=\frac{1}{\sqrt{3}}\left(\begin{array}{ccc}
{1}&{1}&{1}\\
{1}&{\omega}&{\omega^2}\\
{1}&{\omega^2}&{\omega}
\end{array}\right). 
\end{eqnarray*}
Important qutrit gates are given by the permutations $X_{12}=S(2,0,0)$, $X_{02}=XX_{12}X^{\dagger}$ and $X_{01}=XX_{02}X^{\dagger}$, 
\begin{eqnarray}
[X_{12}]_c=\left(\begin{array}{ccc}
{1}&{0}&{0}\\
{0}&{0}&{1}\\
{0}&{1}&{0}
\end{array}\right), \ \ \ 
[X_{02}]_c=\left(\begin{array}{ccc}
{0}&{0}&{1}\\
{0}&{1}&{0}\\
{1}&{0}&{0}
\end{array}\right), \ \ \ [X_{01}]_c=\left(\begin{array}{ccc}
{0}&{1}&{0}\\
{1}&{0}&{0}\\
{0}&{0}&{1}
\end{array}\right).
\end{eqnarray}  
The notation here should be clear: 
$X_{mn}$ corresponds to the restriction of the $X$ operation to the $2-$dimensional subspace spanned by $|m\rangle$ and $|n\rangle$. These permutation gates are sometimes used alternatively in the design of quantum circuits. 

An important qutrit gate that is not an element of the local Clifford group is the so-called qutrit $T$ gate given by
\begin{eqnarray}
[T]_c=\left(\begin{array}{ccc}
{1}&{0}&{0}\\
{0}&{\eta}&{0}\\
{0}&{0}&{\eta^{-1}}
\end{array}\right)
\end{eqnarray}
where $\eta=e^{i2\pi/9}$; this is the gate analogous to the qubit $\pi/8$ gate \cite{quditmagicgates}.

\subsubsection{Multi-controlled unitaries}

Given a bipartite system $\mathcal{H}_a\otimes\mathcal{H}_b$, with orthonormal basis $\{|a,b\rangle\}_{a,b=0}^{d-1}$, 
a controlled unitary $CU$ operation is the bipartite unitary defined as 
\begin{eqnarray}
CU=\sum_{a=0}^{d-1}|a\rangle\langle a|\otimes U^a.
\end{eqnarray}
where $U$ is an unitary acting on $\mathcal{H}_b$. 
The number of times that $U$ is applied on the second subsystem is then conditioned on the control qudit $|a\rangle$. 
We can extend recursively this definition to a tripartite system $\mathcal{H}_a\otimes\mathcal{H}_b\otimes\mathcal{H}_c$ 
with orthonormal basis $\{|a,b,c\rangle\}_{a,b,c=0}^{d-1}$ via 
\begin{eqnarray}
CCU=\sum_{a=0}^{d-1}|a\rangle\langle a|\otimes CU^a
\end{eqnarray}
and proceed by induction in order to obtain the multi-controlled $n-$partite unitary 
$C^{(n)}U=\sum_{a=0}^{d-1}|a\rangle\langle a|\otimes [C^{(n-1)}U]^a$. Moreover, if $\mathcal{I}$ is some index set, we use $C^{(\mathcal{I})}U$ to denote the multi-controlled $U$ gate acting on the qudits labeled by $\mathcal{I}$.

For dimensions $d>2$, there is a division of controlled unitaries into standard controlled unitaries, 
which are those just defined, and the {\it hard controlled} versions of an unitary $U$: 
The hard controlled $|a\rangle\text{-}U$ is defined as
\begin{eqnarray}
|a\rangle\text{-}U= (I-|a\rangle\langle a|)\otimes I +|a\rangle\langle a|\otimes U
\end{eqnarray}
i.e., the unitary is solely conditional on $|a\rangle$. Hard controlled unitaries are fundamental blocks on the construction of multi-controlled unitaries. For example, a qutrit standard $CZ$ gate can be seen as $CZ=(|1\rangle\text{-}Z)(|2\rangle\text{-}Z^2)$.  

Together with the Local Clifford group, the set of controlled unitaries $CU$ acting on $n$ qudits, where $U$ is a Pauli gate, spans the Clifford group on $n$-qudits. The characterization of Clifford and non-Clifford gates is a central problem in quantum information theory, due to results such as the Gottesman-Knill theorem \cite{gkt}, quantum computational models in terms of magic states and gates \cite{braviykitaev} and relations to the foundations of quantum theory \cite{context}.  
Particularly in our work, the Clifford+T set of qutrit gates has the nice property of being approximately universal, in the sense that any qutrit unitary can be arbitrarily approximated by gates from this set \cite{qutrituniv}.

\subsection{Qudit hypergraph states}

A set of states that is closely related to the Clifford group is the set of {\it qudit hypergraph states} \cite{qudithypergraph1,qudithypergraph2}. 
These are the extension to higher-dimensional systems of multi-qubit hypergraph states, where both classes have been used in entanglement theory \cite{hypergraph1,hypergraph2,hypergraph3}, the foundations of quantum theory \cite{hypergraph4} and quantum information \cite{hypergraph5,hypergraph6}. Recent reviews on these topics can be found in \cite{hyperrev1,hyperrev2}.
 
In a given dimension $d$, the idea is to associate to a given multi-hypergraph $H=(V,E)$ a quantum state $|H\rangle$ via the following recipe:
(i) For each vertex $v\in V$, there is a local state $|+_v\rangle=F_v|0_v\rangle$ and we define $|+\rangle^V=\bigotimes_{v\in V}|+_v\rangle$;
(ii) For each hyperedge $e\in E$ with multiplicity $g_e\in\mathbb{Z}_d$, we apply the gate $C^{(e)}Z^{g_e}$ on $|+\rangle^V$. 
Thus the qudit hypergraph state that represents the multi-hypergraph $H=(V,E)$ is given by     
\begin{eqnarray}
|H\rangle=\prod_{e\in E}C^{(e)}Z^{g_e}|+\rangle^V
\end{eqnarray}
The interconvertibility of hypergraph states under the action of the Local Clifford group follows a greatest common divisor hierarchy; alternative definitions and other properties of these states can be found in \cite{qudithypergraph1,qudithypergraph2}. 
Further generalizations of the set of hypergraph states were provided in various works \cite{ffes,hopfcluster,calibrated}, 
with a focus on the mathematical aspects of such classes of states. 
One of our goals is to give an alternative mathematical representation with a physical appeal, 
as well as to present implementations of hypergraph states that are feasible with current experimental techniques.

\section{Angular momentum representation of gates and states}

In what follows, we work with fixed values of $j$ and a given angular momentum basis element $|j;m\rangle$ is denoted simply by $|m\rangle$. We also remark that the terms such as ``quadrupole" and ``multipole" are used loosely. 

\subsection{Local gates}

Let us start by recalling that rotations around the $z$-axis by an angle $\phi$ are given by
\begin{eqnarray}
R_z(\phi)=\exp{i\phi J_z}=\sum_{m=-j}^je^{im\phi}|m\rangle\langle m|
\end{eqnarray}
The matrix representation of this rotation in the ordered angular momentum basis $\{|m=j\rangle,|m=j-1\rangle,\ldots,|m=-j\rangle\}$ is 
\begin{eqnarray*}
[R_z(\phi)]_j=\left(\begin{array}{ccccc}
{e^{i\phi j}}&{}&{}&{}&{}\\
{}&{e^{i\phi (j-1)}}&{}&{}&{}\\
{}&{}&{\ddots}&{}&{}\\
{}&{}&{}&{e^{i\phi (-j+1)}}&{}\\
{}&{}&{}&{}&{e^{i\phi (-j)}}
\end{array}\right)
\end{eqnarray*}
For example, in the case $j=1$ we have
\begin{eqnarray*}
[R_z(\phi)]_j=\left(\begin{array}{ccc}
{e^{i\phi}}&{}&{}\\
{}&{1}&{}\\
{}&{}&{e^{-i\phi}}
\end{array}\right)
\end{eqnarray*}
Setting $\phi=2\pi/3$, we have
\begin{eqnarray*}
[R_z(2\pi/3)]_j=\left(\begin{array}{ccc}
{\omega}&{}&{}\\
{}&{1}&{}\\
{}&{}&{\omega^2}
\end{array}\right)
\end{eqnarray*}
where $\omega=e^{2\pi i/3}$ is the third root of unit. 
Reordering the basis as $\{|m=0\rangle,|m=1\rangle,|m=-1\rangle\}$ and noticing that $(-1)=2$ modulo $3$, 
we can encode the computational basis as $\{|0_L\rangle=|m=0\rangle,|1_L\rangle=|m=1\rangle,|2_L\rangle=|m=-1\rangle\}$ and thus
\begin{eqnarray*}
[R_z(2\pi/3)]_c=\left(\begin{array}{ccc}
{1}&{}&{}\\
{}&{\omega}&{}\\
{}&{}&{\omega^2}
\end{array}\right)=[Z]_c
\end{eqnarray*}   
and we conclude that $R_z(2\pi/3)$ corresponds to the qutrit local gate $Z$. Similarly, $Z^2=R_z(4\pi/3)$.

These observations are easily generalizable to higher dimensions. 
Restricting to $j$ integer, i.e., dimensions $d=2j+1$ odd, we obtain a similar result by encoding our logical qudits as 
$|0_L\rangle=|m=0\rangle,|1_L\rangle=|m=1\rangle,|2_L\rangle=|m=2\rangle,\ldots,|j_L\rangle=|m=j\rangle,
|(j+1)_L\rangle=|m=-j\rangle,|(j+2)_L\rangle=|m=-j+1\rangle,\ldots,|(2j)_L\rangle=|m=-1\rangle$. 
Hence, we are reordering the basis according to the value $m$ modulo $d=2j+1$. 
Under this reordering, it is easy to see that $R_z(2\pi/d)$ corresponds to the qudit local gate $Z$.
Moreover, the other potencies are given by $Z^k=R_z(2k\pi/d)$, $k\in\mathbb{Z}_d$. 

We can relate the discussion to the so-called Pegg-Barnett phase formalism \cite{pg1,pg2}, 
which is one of many attempts to give a consistent definition of phase and time operators in quantum theory. 
The whole conundrum of the quantum phase problem goes way beyond the scope of the present work, 
but we refer the interested reader to \cite{lynch}.  
By definition, the Pegg-Barnett hermitean phase operator is given as $\Theta_z=FJ_zF^{\dagger}$ and  
we have the corresponding gates $X^k=FZ^kF^{\dagger}=\exp[-(2k\pi i/d) \Theta_z]$, $k\in\mathbb{Z}_d$. 
Hence, the angular momentum $J_z$ and phase $\Theta_z$ operators can be seen as generators of the local Pauli group in odd dimensions \cite{vourdas2}. 

For $d$ even, $j$ half-integer case, the situation is not so direct. 
Already in the qubit case $j=1/2$, we have 
\begin{eqnarray}
[R_z(\phi)]_j=[e^{(i\phi/2) \sigma_z}]_j=\left(\begin{array}{cc}
{e^{i\phi/2}}&{0}\\
{0}&{e^{-i\phi/2}}
\end{array}\right)=e^{i\phi/2}\left(\begin{array}{cc}
{1}&{0}\\
{0}&{e^{-i\phi}}
\end{array}\right)
\end{eqnarray}
and, for the construction of multi-controlled rotations, the global phase $e^{i\phi/2}$ has to be taken into account, 
bringing a considerable number of unavoidable extra corrections with increasing number of parties. 
For this reason, we mostly restrict our discussion to odd dimensional systems, 
but stress that the extensions to even dimensional systems are straightforward.

\subsubsection{The $j=1$ case}

In the $j=1$ case we have further interesting simplifications. Noticing that $J_l^3=J_l$ and $J_l^4=J_l^2$, $l=x,y,z$, we see that 
\begin{eqnarray*}
R_l(\phi)&=&\exp(i\phi J_l)=I+i\sin\phi J_l+(\cos\phi-1)J_l^2\\
U_{oat}(\phi,l)&=&\exp(i\phi J_l^2)=I+(e^{i\phi}-1)J^2_l
\end{eqnarray*} 
where $U_{oat}(\phi,l)$ refers to an one-axis-twisting operation \cite{oat1,oat2}. Moreover, an arbitrary Hamiltonian in the $j=1$ case is a polynomial of at most order $2$ in the $J_k$'s; the linear terms are associated to dipole potentials, while the quadratic terms $\{J_k,J_l\}=J_kJ_l+J_lJ_k$ represent quadrupole potentials, 
which are doable in many physical systems.   
In particular,
\begin{eqnarray*}
[R_x(\pi)]_j=\left(\begin{array}{ccc}
{1}&{0}&{0}\\
{0}&{1}&{0}\\
{0}&{0}&{1}
\end{array}\right)-\left(\begin{array}{ccc}
{1}&{0}&{1}\\
{0}&{2}&{0}\\
{1}&{0}&{1}
\end{array}\right)=\left(\begin{array}{ccc}
{0}&{0}&{-1}\\
{0}&{-1}&{0}\\
{-1}&{0}&{0}
\end{array}\right)
\end{eqnarray*}
and reordering gives the local qutrit Clifford gate $[R_x(\pi)]_c=-[X_{12}]_c=-[S(2,0,0)]_c$, up to a $(-1)$ phase. 
As in the even-dimensional case, the $(-1)$ phase has to be taken into consideration, 
but since Clifford gates are usually employed for their group action via conjugation, 
these gates mostly appear in pairs and the $(-1)$ phases cancel out, even in their multi-controlled versions.
However, in other situations this global phase should be taken into account.  

For the $j=1$ case, the Pegg-Barnett phase operator $\Theta_z=FJ_zF^{\dagger}$ can be expressed as
\begin{eqnarray}
[\Theta_z]_j= [F]_j[J_z]_j[F^{\dagger}]_j=\frac{1}{\sqrt{3}}\left(\begin{array}{ccc}
{0}&{i}&{-i}\\
{-i}&{0}&{i}\\
{i}&{-i}&{0}
\end{array}\right)
\end{eqnarray} 
which we readily identify as 
\begin{eqnarray}
\Theta_z= \sqrt{\frac{1}{3}}\{J_y,J_x\}-\sqrt{\frac{2}{3}}J_y
\end{eqnarray} 
or, alternatively, as $\Theta_z=\cos\alpha\{J_y,J_x\}-\sin\alpha J_y$, where $\alpha=\tan^{-1}(\sqrt{2})$ is the so-called magic angle \cite{magicangle}. The interaction given by $\Theta_z$ can be thus be realized by first applying the rotation $R_x(-\alpha)$ to $J_z$, obtaining $\cos\alpha J_z-\sin\alpha J_y$, and then applying an one-axis twisting operation:
\begin{eqnarray}
    e^{-i(\pi/2) J_y^2}(\cos\alpha J_z-\sin\alpha J_y)e^{i(\pi/2) J_y^2}=\Theta_z
\end{eqnarray}
This allow us to obtain the following decomposition for the qutrit DFT :
\begin{eqnarray}
    F=-U_{oat}(-\pi/2,y)R_x(-\alpha)e^{-i(\pi/2)(I-J_z^2)}=U_{oat}(-\pi/2,y)R_x(-\alpha)U_{oat}(-\pi/2,z)e^{i\pi/2}
\end{eqnarray}
We obtain then that $X=\exp[(4\pi i/3)\Theta_z]$ and the remaining local Clifford gates read
\begin{eqnarray}
S(1,1,0)=\exp[(4\pi i/3) J_z^2]; \ \ S(1,2,0)=\exp[(2\pi i/3) J_z^2]; \\ S(1,0,1)=\exp[(2\pi i/3) \Theta_z^2]; \ \ S(1,0,2)=\exp[(4\pi i/3) \Theta_z^2].  
\end{eqnarray}

The qutrit $T$ gate has a simple expression $T=e^{(2\pi i/9) J_z}$, being a simple rotation, which is usually simple to implement. For comparison, in the next odd dimension $d=5$ ($j=2$), we find the expression $\exp[(4\pi i/5)J_z^3]$ \cite{quditmagicgates} for the analogous of this gate, demanding a third-order term that is considerably more demanding in practice.  
Moreover, the qutrit analogous of the qubit $\sqrt{\sigma_z}$ gate is, up to a global phase factor, given by $\exp[(2\pi i/3) (J_z^2-J_z)]$.
In \cite{3clifT}, we employ these expressions to construct alternative experimental realizations of multi-qutrit gates and states.

\subsection{Multi-controlled gates}

A typical interaction in systems described by angular momentum is the following (dipole) coupling: 
\begin{eqnarray*}
J^a_z\otimes J^b_z&=&\left(\sum_{m_a}m_a|m_a\rangle\langle m_a|\right)\otimes\left(\sum_{m_b}m_b|m_b\rangle\langle m_b|\right)\\
&=&\sum_{m_a,m_b}m_am_b|m_a,m_b\rangle\langle m_a,m_b|
\end{eqnarray*}
The unitary operator obtained from exponentiating such coupling is diagonal and given by:
\begin{eqnarray*}
\exp(i\phi J^a_z\otimes J^b_z)&=&\sum_{m_a,m_b}e^{i\phi m_am_b}|m_a,m_b\rangle\langle m_a,m_b|\\
&=&\sum_{m_a}|m_a\rangle\langle m_a|\otimes\left[\sum_{m_b}e^{i\phi m_b}|m_b\rangle\langle m_b|\right]^{m_a}\\
&=& \sum_{m_a}|m_a\rangle\langle m_a|\otimes [R_z(\phi)]^{m_a}
\end{eqnarray*}
i.e., a controlled $z-$rotation. For $j$ integer and $k\in\mathbb{Z}_d$, we have that $\phi=2k\pi/d$ results in
\begin{eqnarray*}
\exp\left[\left(\frac{2k\pi i}{d}\right) J^a_z\otimes J^b_z\right]=\sum_{m_a}|m_a\rangle\langle m_a|\otimes Z^{km_a}
\end{eqnarray*}
and we thus conclude that $CZ^k=\exp[(2k\pi i/d) J^a_z\otimes J^b_z]$. 

For three parties and $j$ integer the ideas are similar and we obtain 
\begin{eqnarray*}
CCZ^k=\exp\left[\left(\frac{2k\pi i}{d}\right) J^a_z\otimes J^b_z\otimes J^c_z\right].
\end{eqnarray*}
The interaction $J^a_z\otimes J^b_z\otimes J^c_z$ is reported in systems composed of spin $1/2$ particles \cite{3spin1,3spin2}, but in  the Appendix we give a possible procedure to implement N-body angular momentum couplings with higher $j$ values in terms of lower ones. 

In an arbitrary n-partite system $\mathcal{H}=\bigotimes_{\nu=1}^n\mathcal{H}_{\nu}$ we have
\begin{eqnarray*}
C^{(n)}Z^k=\exp\left[\frac{2k\pi i}{d}\bigotimes_{\nu=1}^n J^{\nu}_z\right]
\end{eqnarray*}
Hence, for $j$ integer we see that multi-controlled $Z^k$ gates correspond directly to N-body angular momentum couplings between $J_z$ components, 
showing the importance of implementing such type of interactions. 

From the multi-controlled $Z^k$ gates we can obtain other multi-controlled Pauli unitaries via conjugation by suitable local Clifford gates. 
In particular, by applying the local DFT to the last subsystem, we obtain $C^{(n)}X^k=F^{n}[C^{(n)}Z](F^{n})^{\dagger}$. 
In the bipartite case, this reads $CX^k=\exp[-(2k\pi i/d)J_z^a\otimes\Theta_z^b]$ and this gate can be seen as a result of the coupling between 
angular momentum and phase in a discrete phase-space picture.   
Moreover, since $R_x(\pi)Z^ kR_x(-\pi)=(Z^k)^{\dagger}=Z^{-k}$, where $k\in\mathbb{Z}_d$, we have $C^{(n)}Z^{-k}=R^n_x(\pi)[C^{(n)}Z^k]R^n_x(-\pi)$, 
i.e, $C^{(n)}Z^k$ and $C^{(n)}Z^{-k}$ can be mapped into each other simply by a local $x$-axis $\pi$-rotation.  

In the $j=1$ case, since $\Theta_z=\cos\alpha\{J_y,J_x\}-\sin\alpha J_y$, the gate $CX^k=\exp[(4k\pi i/3)J_z^a\otimes\Theta_z^b]$ can be seen as the result of a dipole-quadrupole coupling \cite{dipolequadrupole1,dipolequadrupole2}. 
A similar coupling appears when we consider hard controlled phase gates. Since $|m=\pm 1\rangle\langle m=\pm 1|=(J_z^2\pm J_z)/2$ and $|0\rangle\langle 0|=I-J_z^2$, 
we have $|m\rangle\text{-}Z^k=\exp[(2k\pi i/3) |m_a\rangle\langle m_a|\otimes J_z^b]$. Finally, the hard-control $X$ gates are obtained by $|m\rangle\text{-}X^k=F^b(|m\rangle\text{-}Z^k)(F^b)^{\dagger}=\exp[(4k\pi i/3) |m_a\rangle\langle m_a|\otimes \Theta_z^b]$, corresponding to quadrupole-quadrupole interactions \cite{quadrupolequadrupole}. \newline

{\it Even dimensions} Let us illustrate the differences present in even-dimensional systems by considering the qubit multi-controlled phase gates. 
It is a simple exercise to show the following formula,
\begin{eqnarray}
    C^{(n)}Z=\exp\left[\frac{1}{2^n}(I-\sigma_z)^{\otimes n}\right], \label{formula1}
\end{eqnarray}
which depends not only on the $n$-body interaction $\sigma_z^{\otimes n}$, but also depends on every all-to-all lower-order interactions between $n-1,n-2,\ldots,3,2$ and single qubits. In other words, while the odd-dimensional case requires a single type of interaction, the even-dimensional case demands $2^n$ interactions for each multi-controlled gate. This feature can be traced back to the global phase in the expression $R_z(\pi)=e^{i\pi/2}\left(\begin{array}{cc}{1}&{0}\\{0}&{e^{-i\pi}}\end{array}\right)=i\sigma_z$. If we construct the controlled version of this gate, we obtain
\begin{eqnarray}
    CR_z(\pi)=|0\rangle\langle0|\otimes I+|1\rangle\langle 1|\otimes (i\sigma_z)=|0\rangle\langle0|\otimes I+i|1\rangle\langle 1|\otimes \sigma_z=(\sqrt{\sigma_z}\otimes I)CZ
\end{eqnarray}
and thus $CZ=(\sqrt{\sigma_z}^{\dagger}\otimes I)CR_z(\pi)$. The extra $i$ phase creates a lower-order gate $({\sqrt{\sigma_z}}^{\dagger}\otimes I)$, which needs to be eliminated if we want to generate the gate $CZ$. Each extra qubit demands the elimination of all these residual lower-order gates, which can be a very demanding task. 

    A more general argument can be drawn by extrapolating (\ref{formula1}) to arbitrary even-dimensional systems, via the constructions in \cite{vourdas2}. By a similar reasoning to the odd-dimensional case, we obtain that
\begin{eqnarray}
    C^{(n)}Z^k=\exp\left[\frac{2k\pi i}{d}\left(J_z+\frac{I}{2}\right)^{\otimes n}\right]
\end{eqnarray}
showing once again that angular momentum interactions of all orders are necessary when $d$ is even.

\subsection{Qudit hypergraph states}

As defined previously, the qudit hypergraph state representing a given multi-hypergraph $H=(V,E)$ is given by 
$|H\rangle=\prod_{e\in E}C^{(e)}Z^{g_e}|+\rangle^V$. We have shown in the previous discussion that $C^{(n)}Z^{k}=\exp[(2k\pi /d) J_z^{\otimes n}]$ when $d$ is odd and $ C^{(n)}Z^k=\exp\left[(2k\pi i/d)\left(J_z+I/2\right)^{\otimes n}\right]$ when $d$ is even. For simplicity, we restrict the discussion to odd $d$; we can then write  
\begin{eqnarray}
|H\rangle=\prod_{e\in E}\exp\left[\frac{2g_e\pi i}{d} J^{(e)}_z\right]|+\rangle^V=\exp\left[\frac{2\pi i}{d}\sum_{e\in E}g_e J^{(e)}_z\right]|+\rangle^V
\end{eqnarray}
where $J^{(e)}_z=\bigotimes_{\nu\in e}J^{\nu}_z$. We can then see a qudit hypergraph state as $|H\rangle=e^{(2\pi i/d) G_H}|+\rangle^V$, 
where we define the hypergraph Hamiltonian
\begin{eqnarray}
G_H=\sum_{e\in E}g_e J^{(e)}_z,
\end{eqnarray}
composed of angular momentum interactions between vertices connected by an hyperedge. The relative coupling strengths $g_e$ are numbers in $\mathbb{Z}_d$, which can arise in periodic configurations; alternatively, we can think these numbers as integer multiples of a fundamental coupling strength $g$.
The fiducial state $|+\rangle^V=\bigotimes_{\nu\in V}|+_{\nu}\rangle$ can be seen as the $0$-eigenstate of the collective phase operator $\Theta^{V}_z=\sum_{\nu\in V}\Theta^{\nu}_z$, or as the ground-state of the operator $(\Theta^2_z)^V$. In the $j=1$ we have $\Theta_z^2=-\sqrt{2}J_x+2J_y^2+J_z^2$, which corresponds to the Lipkin-Meshkov-Glick Hamiltonian \cite{lmg1,lmg2}, but for $j>1$ we need higher-order moments of the angular momentum operators, usually a very challenging constraint in practice. In order to circumvent this issue, as well as to see this scheme in a more dynamical fashion, in latter sections we will construct a similar set of states more directly related to the angular momentum framework.

\subsection{Decompositions in terms of bipartite and local gates}

We have shown in previous sections how to realize multi-controlled gates in terms of N-body angular momentum interactions. 
However, both theoretical models and experimental observations tend to deal with two-body and/or local Hamiltonians. 
In the angular momentum representation, terms such as $\mathbf{S}_a\cdot\mathbf{S}_b$ or $\mathbf{L}_a\cdot\mathbf{S}_b$ are commonplace, 
while explicit three-body or more interactions are relatively more recent in the literature. Moreover, in our discussions one example of expression is the qutrit  gate $CCX=\exp[(4\pi i /3)J_z\otimes J_z\otimes \Theta_z]$. In a few-body regime, it is possible to argue that $CCX=(I\otimes I\otimes F)CCZ(I\otimes I\otimes F^{\dagger})$; however, in a many-body scenario, the allowed operations are collective, thus requiring a native  dipole-dipole-quadrupole interaction, and there is currently no realistic situation where this type of coupling occurs. For $j>1$ the situation is even more complex, since many multipole interactions of different orders would be needed. We thus invoke results concerning the decomposition of multi-controlled gates as sequences of bipartite and local gates, which demand two-body couplings at most. In the next section we map these constructions into a quantum optical realization, bringing our scheme closer to current available technology. We leave open the important questions about optimization or non-idealities \cite{noise1,noise2} for future investigations. 

For the $j=1$ qutrit case, we invoke Lemma 1 of \cite{qutritgates1}, a circuit decomposing a three-qutrit hard-controlled Pauli gate $|q\rangle\text{-}CW$, $q=0,1,2$ ($W^3=W$), in terms of two-qutrit hard-controlled Pauli gates:
\begin{eqnarray}
    |q\rangle\text{-}CW=(|q_a\rangle\text{-}W_c^2)(|q_a\rangle\text{-}X_b)(|1_b\rangle\text{-}W_c)(|q_a\rangle\text{-}X_b)(|1_b\rangle\text{-}W_c^2), \ q=0,1,2.
\end{eqnarray}
Here the subscripts $a,b,c$ label the subsystems of the three-qutrit state space $\mathcal{H}_a\otimes\mathcal{H}_b\otimes\mathcal{H}_c$. We obtain then the following decompositions:  
\begin{eqnarray*} 
|1\rangle\text{-}CZ&=&(|1_b\rangle\text{-}Z_c^2)(|1_a\rangle\text{-}X_b^2)(|1_b\rangle\text{-}Z_c)(|1_a\rangle\text{-}X_b)(|1_a\rangle\text{-}Z_c^2)\\
|2\rangle\text{-}CZ^2&=&(|1_b\rangle\text{-}Z_c)(|2_a\rangle\text{-}X_b^2)(|1_b\rangle\text{-}Z^2_c)(|2_a\rangle\text{-}X_b)(|2_a\rangle\text{-}Z_c)
\end{eqnarray*}
Thus we get the full $CCZ=(|1\rangle\text{-}CZ)(|2\rangle\text{-}CZ^2)$ by performing these operations sequentially - in any order, since the block gates $|1\rangle\text{-}CZ$ and $|2\rangle\text{-}CZ^2$ commute.

Conjugation by the local DFT on the target qutrit gives us the gates $|1\rangle\text{-}CX$, $|2\rangle\text{-}CX^2$ and $CCX$. 
Moreover, we have that $CCZ^2=R^c_x(\pi)[CCZ]R^c_x(-\pi)$ and $CCX^2=R^c_x(\pi)[CCX]R^c_x(-\pi)$.  
Furthermore, from the controlled Pauli gates we can implement controlled Clifford gates by going up in the Clifford hierarchy, e.g., 
by conjugating a given controlled Pauli gate with a suitable non-Clifford unitary such as the $T$ gate.   
In \cite{qutritgates1} one finds various decompositions of qutrit multi-controlled Clifford+T gates using lower-order gates, with possible additions of ancillas. More generally, one can employ the techniques of \cite{quditgates} for the construction of multi-controlled qudit gates.

\section{Quantum optical analog}

Using the results of the previous sections, we build a quantum optical analog of the qutrit $CCZ$ gate and of an uniform qutrit hypergraph state. 
We focus on optical setups, but the interactions and operations described here can be easily translated to other multimode bosonic systems such as circuit QED \cite{cqed,cqed2} and Bose-Einstein condensates \cite{bec}. 

The seminal work \cite{multirail1} gives a method to construct any $d\times d$ unitary by the sole use of linear optical operations and single-photon sources; see as well \cite{multirail2}. These results enable us to construct any desired local qudit gate simply via linear couplings between bosonic modes and phase-shifts. Compared with the alternative scheme in \cite{3clifT}, or even the angular momentum realizations of the previous sections, both restricted to the qutrit case and requiring nonlinear interactions even for the local gates, the present proposal has a clear advantage. The tradeoff here is the increase in the number of modes for increasing local dimensions: a $N$-qudit system requires $N.d$ modes. 

In the general case, we consider a system with $d$ modes $\mathcal{H}=\bigotimes_{\nu=0}^{d-1}\mathcal{H}_{\nu}$, 
with creation and annihilation operators $a_{\nu}$, $a^{\dagger}_{\nu}$ acting on $\mathcal{H}_{\nu}$ satisfying canonical commutation relations
\begin{eqnarray}
[a_{\nu},a^{\dagger}_{\nu'}]=\delta_{\nu\nu'}I; \ \ [a_{\nu},a_{\nu'}]=[a^{\dagger}_{\nu},a^{\dagger}_{\nu'}]=0.
\end{eqnarray}
We employ the multi-rail encoding $|q_L\rangle=a_q^{\dagger}|0,\ldots,0\rangle$, $q=0,1,\ldots,d-1$, or explicitly: 
$|0_L\rangle=|1,0,\ldots,0\rangle, |1_L\rangle=|0,1,0,\ldots,0\rangle,\ldots,|(d-1)_L\rangle=|0,\ldots,0,1\rangle$.

Specializing to a qutrit, 
we have $|0_L\rangle=a_0^{\dagger}|0,0,0\rangle=|1,0,0\rangle, |1_L\rangle=a_1^{\dagger}|0,0,0\rangle=|0,1,0\rangle,|2_L\rangle=a_2^{\dagger}|0,0,0\rangle=|0,0,1\rangle$. 
The three-mode Jordan-Schwinger map of the angular momentum operators \cite{multijsm} gives
\begin{eqnarray*}
J_x&=&\frac{1}{\sqrt{2}}[a_0^{\dagger}(a_1+a_2)+a_0(a_1^{\dagger}+a_2^{\dagger})], \\  J_y&=&\frac{i}{\sqrt{2}}[a_0^{\dagger}(a_1+a_2)-a_0(a_1^{\dagger}+a_2^{\dagger})], \\  J_z&=&a_1^{\dagger}a_1-a_2^{\dagger}a_2,
\end{eqnarray*}
as can be readily verified from their commutation relations. An important feature here is the conservation of the total number $N=a_0^{\dagger}a_0+a_1^{\dagger}a_1+a_2^{\dagger}a_2$. 
The operator $J_z$ is simply the population difference between modes $1$ and $2$, while the operators 
$J_x$ and $J_y$ represent sequential beam splitter interactions that differ only due to local phases, which we maintain implicit. 

One key ingredient for the implementation of controlled gates in our optical scheme is the two-mode cross-Kerr interaction \cite{boyd}, which is given by
\begin{eqnarray}
    H_{ck}=\chi a_{\nu}^{\dagger}a_{\nu}a_{\mu}^{\dagger}a_{\mu}
\end{eqnarray}
where $\chi$ is the interaction strength. In the optical context, $\chi$ has typically small values and this attribute tends to cast doubts on the feasibility of the cross-Kerr interaction for gate realizations \cite{shapiro}. However, there are proposals that address this issue \cite{ckphase,ckphase2,mmck2}. In particular, the schemes in \cite{ckphase,ckphase2} are more suited for our encoding, since they preserve the total number of photons, while \cite{mmck2} is based on squeezing interactions, introducing higher photon excitations and consequently a certain degree of non-determinism. Alternatively, the phases required in our formulation are integer multiples of $2\pi/d$; for high values of $d$, the required interaction strength would then be lower. However, as already noted, higher $d$ implies a higher number of modes and we are, in this sense, in a situation similar to the depth-width dichotomy in computer science. In superconducting circuits, the cross-Kerr strength is considerably higher and tunable, but the required low temperatures and usual microwave regime impose scalability limitations. In the context of coupled Bose-Einstein condensates \cite{beccc,beccc2}, an analogous interaction can be considered; the tradeoff in this scenario is between the interaction strength and the stability itself of the condensate. 

Hence, for the sake of the argument and considering the proposals in \cite{ckphase,ckphase2,mmck2}, we assume that the cross-Kerr interaction can attain high enough values in an optical scenario. The basic idea then is to apply the Jordan-Schwinger map on interactions of the form $J_z^{\nu}J_z^{\mu}$, obtaining
\begin{eqnarray}
    J_z^{\nu}J_z^{\mu}&=&[(a_1^{\nu})^{\dagger}a_1^{\nu}-(a_2^{\nu})^{\dagger}a_2^{\nu}][(a_1^{\mu})^{\dagger}a_1^{\mu}-(a_2^{\mu})^{\dagger}a_2^{\mu}]\\
    &=&(a_1^{\nu})^{\dagger}a_1^{\nu}(a_1^{\mu})^{\dagger}a_1^{\mu}-(a_2^{\nu})^{\dagger}a_2^{\nu}(a_1^{\mu})^{\dagger}a_1^{\mu}-(a_1^{\nu})^{\dagger}a_1^{\nu}(a_2^{\mu})^{\dagger}a_2^{\mu}+(a_2^{\nu})^{\dagger}a_2^{\nu}(a_2^{\mu})^{\dagger}a_2^{\mu}\\
    &=&\chi^{-1}(H_{ck}^{1,1}-H_{ck}^{2,1}-H_{ck}^{1,2}+H_{ck}^{2,2})
\end{eqnarray}
A similar reasoning can be used for $N$-body angular momentum interactions $J_z^{\otimes N}$; multi-mode versions of the cross-Kerr interaction have been investigated in the literature \cite{3body,mmck1,mmck2} and could potentially save a considerable number of resources, when considering the implementation of multi-controlled gates.

\subsection{Local gates}

The three-mode Jordan-Schwinger map implies that $J_z=a_1^{\dagger}a_1-a_2^{\dagger}a_2$; by previous discussions, we have $Z=e^{(2\pi i/3)J_z}=e^{(2\pi i/3)a_1^{\dagger}a_1}e^{(4\pi i/3)a_2^{\dagger}a_2}$ and $Z^2=e^{(4\pi i/3)J_z}=e^{(4\pi i/3)a_1^{\dagger}a_1}e^{(2\pi i/3)a_2^{\dagger}a_2}$, which are phase shift operations on modes $1$ and $2$. 

Following the procedure in \cite{multirail1}, the local qutrit DFT can be decomposed as the product $F=U_1U_2U_3U_4$ of the following gates 
\begin{eqnarray}
U_1=\left(\begin{array}{ccc}
{1}&{0}&{0}\\
{0}&{1/\sqrt{2}}&{-1/\sqrt{2}}\\
{0}&{1/\sqrt{2}}&{1/\sqrt{2}}
\end{array}\right), \ U_2=\left(\begin{array}{ccc}
{1}&{0}&{0}\\
{0}&{1}&{0}\\
{0}&{0}&{e^{i\pi/2}}
\end{array}\right), \\ U_3=\left(\begin{array}{ccc}
{1/\sqrt{3}}&{-\sqrt{2/3}}&{0}\\
{\sqrt{2/3}}&{1/\sqrt{3}}&{0}\\
{0}&{0}&{1}
\end{array}\right), \ U_4=\left(\begin{array}{ccc}
{1}&{0}&{0}\\
{0}&{-1/\sqrt{2}}&{-1/\sqrt{2}}\\
{0}&{-1/\sqrt{2}}&{1/\sqrt{2}}
\end{array}\right)
\end{eqnarray}
The gates $U_1$ and $U_4$ are recognized as balanced beam splitting interactions between modes $1$ and $2$,
\begin{eqnarray}
U_1=\exp[-i\pi/4(-ia^{\dagger}_1a_2+ia_1a^{\dagger}_2)], \ \ U_4=U_1e^{i\pi a^{\dagger}_1a_1},
\end{eqnarray}
while $U_2=e^{(\pi i/2) a^{\dagger}_2a_2}$ is a simple phase-shift on mode $2$ and 
\begin{eqnarray}
U_3=\exp[-i\alpha(-ia^{\dagger}_0a_1+ia_0a^{\dagger}_1)]
\end{eqnarray}
represents an unbalanced beam splitter interaction between modes $0$ and $1$, where $\alpha=\tan^{-1}(\sqrt{2})$ denotes the magic angle of previous sections. 
The optical implementation of $F$ is depicted bellow; alternative schemes can be found in \cite{demirel,cilluffo}. Conjugation by $F$ yields the gates $X=FZF^{\dagger}$ and $X^2=FZ^2F^{\dagger}$. 

 \begin{figure}[h]\centering\includegraphics[scale=0.7]{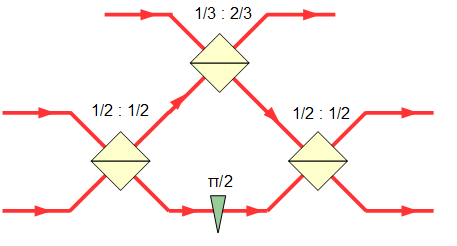}
   \caption{Diagram for the realization of the qutrit gate $F$. }
    \end{figure}

 \begin{figure}[h]\centering\includegraphics[scale=0.6]{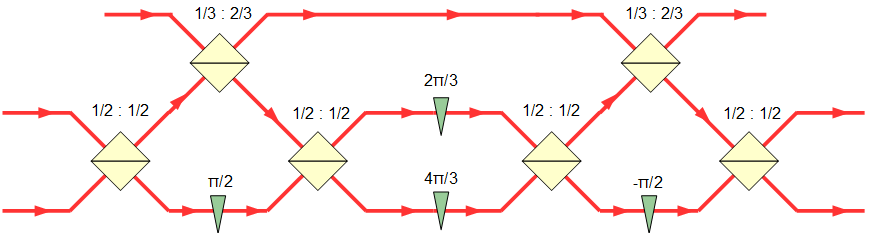}
   \caption{Diagram for the realization of the qutrit gate $X$. }
    \end{figure}

The local Clifford gates $S(1,1,0)=|0\rangle\langle 0|+\omega^2|1\rangle\langle 1|+\omega^2|2\rangle\langle 2|$ 
and $S(1,2,0)=|0\rangle\langle 0|+\omega|1\rangle\langle 1|+\omega|2\rangle\langle 2|$ are diagonal in the computational basis and we can 
implement each via phases shifts on the local modes $1$ and $2$:
\begin{eqnarray}
S(1,1,0)=e^{(4\pi i/3) a^{\dagger}_1a_1}e^{(4\pi i/3)a^{\dagger}_2a_2}; \ \ \ S(1,2,0)=e^{(2\pi i/3) a^{\dagger}_1a_1}e^{(2\pi i/3) a^{\dagger}_2a_2}.
\end{eqnarray}
Similarly, the gates $S(1,0,1)=F(|0\rangle\langle 0|+\omega|1\rangle\langle 1|+\omega|2\rangle\langle 2|)F^{\dagger}$ 
and $S(1,0,2)=F(|0\rangle\langle 0|+\omega^2|1\rangle\langle 1|+\omega^2|2\rangle\langle 2|)F^{\dagger}$ are obtained by 
conjugation by the DFT on appropriate phase-shifts. 

Finally, we have $S(2,0,0)=X_{12}=-\exp[i\pi J_x]$, composed of sequential beam splitting operations. Since the total number $a_0^{\dagger}a_0+a_1^{\dagger}a_1+a_2^{\dagger}a_2$ is conserved, one can apply a collective phase shift $e^{i\pi a_0^{\dagger}a_0}e^{i\pi a_1^{\dagger}a_1}e^{i\pi a_2^{\dagger}a_2}$, in order to eliminate the global $(-1)$ phase. 

 \begin{figure}[h]\centering\includegraphics[scale=0.7]{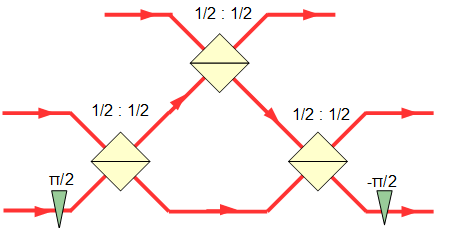}
   \caption{Diagram for the realization of the qutrit gate $-X_{12}=-X_{12}^{\dagger}$. }
    \end{figure}

The qutrit gate $T=e^{(2\pi i/9)J_z}$ is composed simply of phase shifting operations on modes $1$ and $2$, i.e., $T=e^{(2\pi i/9)a_1^{\dagger}a_1}e^{(7\pi i/9)a_2^{\dagger}a_2}$.

\subsection{Multi-controlled unitaries}

We consider bipartite and tripartite qutrit controlled unitaries. We divide each subsystem into three sets $\{a_0,a_1,a_2\}$, $\{b_0,b_1,b_2\}$ and $\{c_0,c_1,c_2\}$, where the labels $0,1,2$ represent the local levels. We have thus nine modes in total, with creation and annihilation operators satisfying the commutation relations
\begin{eqnarray}
    [a_{\mu},a^{\dagger}_{\nu}]=[b_{\mu},b^{\dagger}_{\nu}]=[c_{\mu},c^{\dagger}_{\nu}]=\delta_{\mu\nu}I
\end{eqnarray}
and the remaining commutators are null. The basis of the global state space is given by
\begin{eqnarray}
    |p_{a},q_{b},r_{c}\rangle\equiv a^{\dagger}_pb^{\dagger}_qc^{\dagger}_r|\text{vac}\rangle, \ \ p,q,r=0,1,2,
\end{eqnarray}
where the vacuum state is $|\text{vac}\rangle=|0\rangle^{\otimes 9}$. The two-mode cross-Kerr interaction between $a_k$ and $b_l$ is given by 
\begin{eqnarray*}
H^{a_k,b_l}_{ck}=\chi a_k^{\dagger}a_kb_l^{\dagger}b_l
\end{eqnarray*}
and, as shown previously, we have
\begin{eqnarray*}
J^{a}_zJ^{b}_z=\chi^{-1}(H^{a_1,b_1}_{ck}-H^{a_1,b_2}_{ck}-H^{a_2,b_1}_{ck}+H^{a_2,b_2}_{ck})
\end{eqnarray*}
The four terms above all commute and, recalling the relations $CZ=\exp\left[(2\pi i/3) J^{a}_zJ^{b}_z\right]$ and $CZ^2=\exp\left[(4\pi i/3) J^{a}_zJ^{b}_z\right]$, we have
\begin{eqnarray*}
CZ&=&\exp\left[\frac{2\pi i}{3\chi} H^{a_1,b_1}_{ck}\right]\exp\left[\frac{4\pi i}{3\chi} H^{a_2,b_1}_{ck}\right]\exp\left[\frac{4\pi i}{3\chi} H^{a_1,b_2}_{ck}\right]\exp\left[\frac{2\pi i}{3\chi} H^{a_2,b_2}_{ck}\right]\\
CZ^2&=&\exp\left[\frac{4\pi i}{3\chi} H^{a_1,b_1}_{ck}\right]\exp\left[\frac{2\pi i}{3\chi} H^{a_2,b_1}_{ck}\right]\exp\left[\frac{2\pi i}{3\chi} H^{a_1,b_2}_{ck}\right]\exp\left[\frac{4\pi i}{3\chi} H^{a_2,b_2}_{ck}\right]
\end{eqnarray*}

Since on each subsystem the maximum number of photons is $1$, the number operator on each mode corresponds to the projector onto the single-photon state. We can thus implement the hard-controlled phase gates as
\begin{eqnarray*}
|1\rangle\text{-}Z &=& \exp\left[\frac{2\pi i}{3} a_1^{\dagger}a_1J^b_z\right]=\exp\left[\frac{2\pi i}{3\chi} H^{a_1,b_1}_{CK}\right]\exp\left[\frac{4\pi i}{3\chi}H^{a_1,b_2}_{CK}\right]\\
|2\rangle\text{-}Z &=& \exp\left[\frac{2\pi i}{3} a_2^{\dagger}a_2J^b_z\right]=\exp\left[\frac{2\pi i}{3\chi} H^{a_2,b_1}_{CK}\right]\exp\left[\frac{4\pi i}{3\chi}H^{a_2,b_2}_{CK}\right]\\
|1\rangle\text{-}Z^2 &=& \exp\left[\frac{4\pi i}{3} a_1^{\dagger}a_1J^b_z\right]=\exp\left[\frac{4\pi i}{3\chi} H^{a_1,b_1}_{CK}\right]\exp\left[\frac{2\pi i}{3\chi}H^{a_1,b_2}_{CK}\right]\\
|2\rangle\text{-}Z^2 &=& \exp\left[\frac{4\pi i}{3} a_2^{\dagger}a_2J^b_z\right]=\exp\left[\frac{4\pi i}{3\chi} H^{a_2,b_1}_{CK}\right]\exp\left[\frac{2\pi i}{3\chi}H^{a_2,b_2}_{CK}\right]
\end{eqnarray*}
The corresponding hard-controlled gates $|q\rangle\text{-}X^k$ - $k,q\in \{1,2\}$ - can be constructed via conjugation by the appropriate local Clifford gates. As discussed in the previous section, with these gates we can construct the various three-qutrit gates $|q\rangle\text{-}CZ^k$, $|q\rangle\text{-}CX^k$, $CCZ^k$ and $CCX^k$.  
 \begin{figure}[h]\centering\includegraphics[scale=0.5]{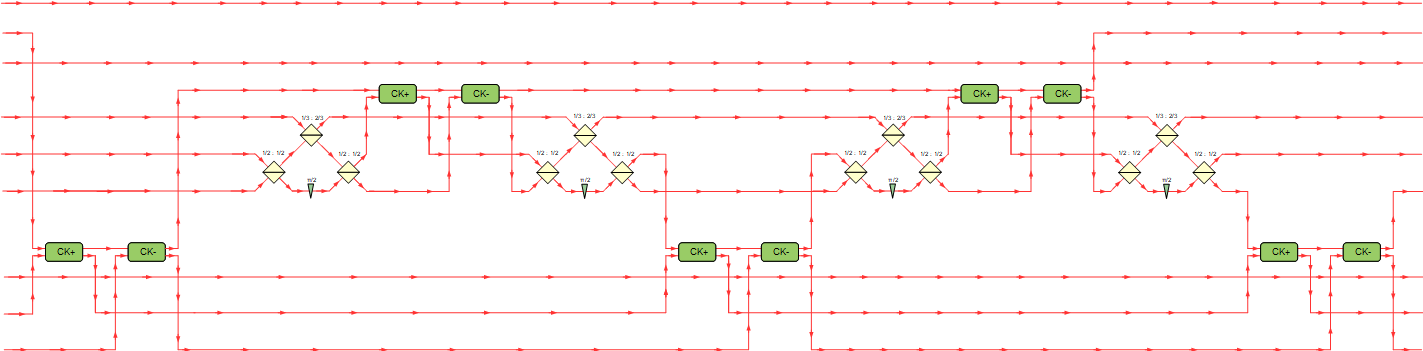}
    \caption{Diagram for the implementation of the three-qutrit hard-controlled phase gate $|1\rangle\text{-}CZ$. }
    \end{figure}

\subsection{Implementation of three-qutrit hypergraph states}

The uniform three-qutrit hypergraph state is given by $|H\rangle = CCZ|+_a,+_b,+_c\rangle$. 
The optical scheme to implement the gate $CCZ$ was described in the previous subsection, 
thus it remains to implement the vertex state $|+\rangle=(1/\sqrt{3})(|0_L\rangle+|1_L\rangle+|2_L\rangle)=(1/\sqrt{3})(|1,0,0\rangle+|0,1,0\rangle+|0,0,1\rangle)$ on each subsystem. Since this is an equal superposition, it can be obtained in an obvious way by sending the single-photon state $|1,0,0\rangle$ through a balanced three-port beam-splitter. Interestingly, from the DFT decomposition, this is equivalent to $U_1U_3|1,0,0\rangle$. 

The other qutrit hypergraph states can be obtained by suitable applications of the local Clifford gates $X_{12}$, $X$ or $Z$ \cite{qudithypergraph1}, and the optical setups implementing these gates were described in previous subsections. 

\section{Angular momentum hypergraph states}

As developed previously, to the multihypergraph $H=(V,E)$ there corresponds the qudit hypergraph state 
\begin{eqnarray}
|H\rangle=\prod_{e\in E}C^{(e)}Z^{g_e}|+\rangle^V=\exp\left[\frac{2\pi i}{d}\sum_{e\in E}g_eJ^{(e)}_z\right]|+\rangle^V
\end{eqnarray} 
In odd dimensions, each multi-controlled gate $C^{(e)}Z^{g_e}$ is produced by angular momentum couplings $J^{(e)}_z=\bigotimes_{k\in e} J_z^k$, 
suggesting a direct relation between models where N-body angular momentum interactions are present and qudit hypergraph states. 
However, in general this connection is not practical, since the vertice state $|+\rangle=F|m=0\rangle$ 
is an eigenstate of the Pegg-Barnett phase operator $\Theta_z$ and not of any angular momentum component $J_l$, $l=x,y,z$. 
Only in special cases such as $j=1$ can we establish a satisfactory correspondence and even then we need to consider quadratic terms for the $J_l$'s.
Moreover, for integer $j$ the $|m=0\rangle$ state is not a $SU(2)$ coherent state, which may hinder its practical implementation. 

In order to address these issues, we define a class of states that is a modification of the qudit hypergraph class, 
but is described solely in terms of N-body angular momentum couplings and eigenstates of the angular momentum operators. 
We call these states {\it angular momentum hypergraph states}; in the special cases where the operators represent spin, 
we denote these states as {\it spin hypergraph states}.
\begin{definition}
Given a weighted hypergraph $H=(V,E)$, we associate the so-called angular momentum hypergraph state via
\begin{eqnarray}
|J_H\rangle=\exp[-i\sum_{e\in E}\phi_e J^{(e)}_z]|x+\rangle^V
\end{eqnarray}
where $\phi_e\in\mathbb{R}$ are the weights of each hyperedge $e\in E$ and $|x+\rangle$ is the $J_x$ eigenstate with the highest eigenvalue (weight). 
\end{definition}
An extra advantage of this definition is the relaxation on the weights $\phi_e$, in order to allow N-body interactions of any strength; see, e.g., \cite{univwgs}. Moreover, since these values contain the time variable implicitly, we can picture the evolution of the state in terms of the relative hyperedges' weights in a given instant.  

One interesting property that follows from the above definition is the following:
\begin{observation}
    Given the qudit hypergraph state  $|H\rangle$, associated to the multi-hypergraph $H=(V,E)$, there exists local invertible operations $\bigotimes_{v\in V}A_v$ such that $\bigotimes_{v\in V}A_v|H\rangle=|J_H\rangle$, where $|J_H\rangle$ is an angular momentum hypergraph state associated to the same multi-hypergraph $H=(V,E)$.   
\end{observation}
{\it Proof:} Each vertex state $|x+\rangle$ has the following expansion in the angular momentum basis: 
\begin{eqnarray}
|x+\rangle=\frac{1}{2^{j}}\sum_{m=j}^{-j}\sqrt{\frac{(2j)!} {(j-m)!(j+m)!}}|m\rangle
\end{eqnarray}
Let $A_v=\sum_m\sqrt{\frac{d}{2^j}\frac{(2j)!} {(j-m)!(j+m)!}}|m\rangle\langle m|$; then we have that $\bigotimes_{v\in V}A_v|+\rangle^V=|x+\rangle^ V$. Moreover, the operator $\bigotimes_{v\in V}A_v$ is diagonal and thus commutes with $\prod_{e\in E}C^{(e)}Z^{g_e}$. We conclude that $\bigotimes_{v\in V}A_v|H\rangle=|J_H\rangle$. $\square$  

When two pure states can be mapped into each other by invertible local operations, we say that they are equivalent by stochastic local operations and classical communication (SLOCC), or SLOCC-equivalent. The classification of multipartite states in terms of their SLOCC-equivalence classes is a central problem in quantum information theory; see, for example, \cite{elos} and references therein.   

A simple property of an angular momentum hypergraph state is that the collective conjugation of $R_x(\pi)^{\otimes n}$ or $R_y(\pi)^{\otimes n}$ 
on $J_z^{\otimes n}$ changes the sign of this term if $n$ is odd and 
maintains the sign if $n$ is even; 
similarly, the action of $R_z(\pi)$ or $R_y(\pi)$ on $|x+\rangle$ maps it to $|x-\rangle$. 
Hence, different combinations of local collective $SU(2)$ actions possibly result in novel symmetries and selection rules. 
These and other properties of angular momentum hypergraph states will be explored in future works.  

\section{Conclusions and perspectives}

In the present work we established a direct connection between qudit multi-controlled gates and N-body angular momentum couplings. We identified the relevant interactions that enable the implementation and representation of gates in the Clifford+T set, with a detailed account of the qutrit case. We believe our work can motivate a search for the occurrence of these types of interactions in atomic physics, condensed matter or quantum optics, or novel methods for their synthesis in experiments.  

Moreover, we applied the angular momentum representation to the characterization of qudit hypergraph states, showing the advantages and drawbacks of this framework. Introducing the set of angular momentum hypergraph states, we managed to circumvent some of the obstacles for practical realizations. For future works, we aim to apply this new class of states to problems that involve quantum N-body physics. 

Finally, we gave a proof of concept of our approach by devising a quantum optical analog of the angular momentum implementation. The scheme obtained demands solely single-photon sources, linear optical processes and cross-Kerr nonlinearities, these being well-known elements of quantum optics. 

Our work seems to illustrate a remarkable but overlooked feature of quantum information theory: by seeking solutions to quantum computational problems, many new physical insights are obtained.  

\section{Acknowledgments}

The author is thankful to Thiago Tunes, Thiago Toledo, Vitor A. Garcia, Gilberto Brito, Daniel Felinto, Daniel Brod, Pierre Louis, Marcos C. de Oliveira, Alexandre Ribeiro, Eduardo Duzzioni, Gustavo Cunha, Helena Bran\c ca and Alexandre Dodonov for ideas and recommendation of references. This work is part of the institutional project ``Mathematical aspects of quantum entanglement" from Universidade Federal de Mato Grosso and was partially supported by the brazilian agency FINATEC. 

\section*{Appendix - Strategy based on the sum of angular momenta}

We obtain a simple procedure to construct N-body angular momentum interactions for systems composed of particles having $j>1/2$ from systems with $j=1/2$ particles. A similar approach can be found in the recent work \cite{tcm}, in terms of the Tavis-Cummings model. For simplicity, we restrict the discussion to interactions of the form $J_z^{\otimes n}$, but the ideas are easily generalizable to other angular momentum interactions. 

It is well-known \cite{sakurai} that the operator $J_z$, for total angular momentum $j$, can be expressed as the sum $J_z=(1/2)\sum_{k=1}^{2j}\sigma_z^k$, where $\sigma_z^1=\sigma_z\otimes I\otimes\ldots\otimes I,\sigma_z^2=I\otimes\sigma_z\otimes I\otimes\ldots\otimes I,\ldots,\sigma_z^{2j}=I\otimes\ldots\otimes I\otimes\sigma_z$. We thus obtain  
\begin{eqnarray}
    \bigotimes_{\nu=1}^nJ_z^{\nu}=\frac{1}{2^n}\bigotimes_{\nu=1}^n\left(\sum_{k=1}^{2j}\sigma_z^{\nu,k}\right)=\frac{1}{2^n}\left(\sum_{k=1}^{2j}\bigotimes_{\nu=1}^n\sigma_z^{\nu,k}\right),
\end{eqnarray}
and we conclude that interactions of the form $J_z^{\otimes n}$ can be constructed from $2^n$ multi-qubit interactions of the form $\sigma_z^{\otimes n}$. Hence, if an experimental setup can implement tunable multi-qubit interactions, in principle it is possible to obtain higher angular momentum interactions of the same order. Since $j=1/2$ couplings are dipolar, there is an interesting advantage in this approach. However, the cost is analogous to the multi-rail encoding described in the main text, in that the number of sites is increased and thus their distribution in three-dimensional space has to be considered. 


 \begin{figure}[h]\centering\includegraphics[scale=0.5]{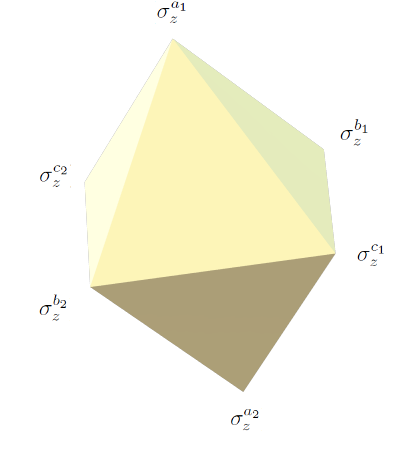}
   \caption{Schematic representation of multiple three-qubit interactions in an octahedral disposition ($j=1$ case).  }
    \end{figure}

Let us illustrate with the $j=1$ three-body interaction $J^a_zJ^b_zJ^c_z$. In this case, in each subsystem we have $J_z=(1/2)(\sigma_z^1+\sigma_z^2)$ and we obtain
\begin{eqnarray}
    J^a_zJ^b_zJ^c_z&=&\frac{1}{8}(\sigma_z^{a_1}+\sigma_z^{a_2})(\sigma_z^{b_1}+\sigma_z^{b_2})(\sigma_z^{c_1}+\sigma_z^{c_2})\\
    &=&\frac{1}{8}\sum_{i,j,k=1}^2\sigma_z^{a_i}\sigma_z^{b_j}\sigma_z^{c_k}
\end{eqnarray}
We can arrange the six points $a_1,a_2,b_1,b_2,c_1,c_2$ as vertices of an octahedron, where the faces represent the three-qubit interactions $\sigma_z^{a_i}\sigma_z^{b_j}\sigma_z^{c_k}$.

\end{document}